\documentclass[aps,prl,showpacs,floats,twocolumn,amsfonts,amssymb,unsortedaddress,floatfix]{revtex4}
\input epsf
\usepackage{bm}
\usepackage{amsmath}
\usepackage{graphicx}

\begin{document}

\renewcommand{\bottomfraction}{0.7}
\renewcommand{\topfraction}{0.7}
\renewcommand{\textfraction}{0.2}
\renewcommand{\floatpagefraction}{0.7}
\renewcommand{\thesection}{\arabic{section}}

\addtolength{\topmargin}{10pt}


\def\Bbb{\mathbb}
\def\d{\partial}
\def\({\left(}
\def\){\right)}

\title{Carbon Nanostructures as an Electromechanical Bicontinuum}
\author{Cristiano Nisoli$^{*}$, Paul E. Lammert$^{*}$, Eric Mockensturm$^{\dagger}$  and Vincent H. Crespi$^{*}$}
\affiliation{$^{*}$Department of Physics and Materials Research Institute \\
$^{\dagger}$Department of Mechanical and Nuclear Engineering\\ The
  Pennsylvania State University, University Park, PA 16802-6300}

\date{\today}
\begin{abstract}
A two-field model provides an unifying framework for elasticity, lattice dynamics and electromechanical coupling in graphene and carbon nanotubes, describes optical phonons, nontrivial acoustic branches, strain-induced gap opening, gap-induced phonon softening, doping-induced deformations, and even the hexagonal graphenic Brillouin zone, and thus  explains and extends a previously disparate accumulation of analytical and computational results. 
 \end{abstract}


\maketitle

Vibrations in carbon nanostructures such as tubes, fullerenes, or graphene sheets~\cite{ji, imperial, Dresselhaus} have a ubiquitous influence on electronic, optical and thermal response: scattering from optical phonons limits charge transport in otherwise ballistic nanotube conductors~\cite{Tans, Kane}; twist deformations  gap metallic tubes~\cite{Rochefort,Yang}; ballistic phonons transport heat in nanotubes with great efficiency~\cite{Berber,Chiu,Kim}; resonant Raman spectroscopy can unambiguously identify a tube's wrapping indices (n,m)~\cite{Richter, Saito, Jorio, Jorio2}; electron-phonon interactions may ultimately limit the electrical performance of graphene~\cite{Novoselov,Zhou}.  Computationally intensive atomistic models of lattice dynamics often lack simplified model descriptions that can facilitate insight, yet traditional analytical continuum models~\cite{ji,Mahan1,Suzu,imperial}, while very  useful and important, cannot describe atomistic phenomena without phenomenological extensions~\cite{Comas,Chico,Gartstein}. Although continuum models are restricted to long-wavelength physics, they have been used to describe  atomic-scale phenomena in bulk binary compounds by incorporating a separate continuum field for each 
sublattice~\cite{mindlin}: in graphene, two fields are necessary. Here we present an analytical ``bicontinuum'' model that represents the full atomistic detail of the graphenic lattice, including optical modes, nonlinear dispersion of in-plane phonons, electromechanical effects and even the hexagonal graphenic Brillouin zone, a construct generally held to be exclusively atomistic.

\begin{figure}[t!]
\epsfxsize 1.5 in
\epsfbox{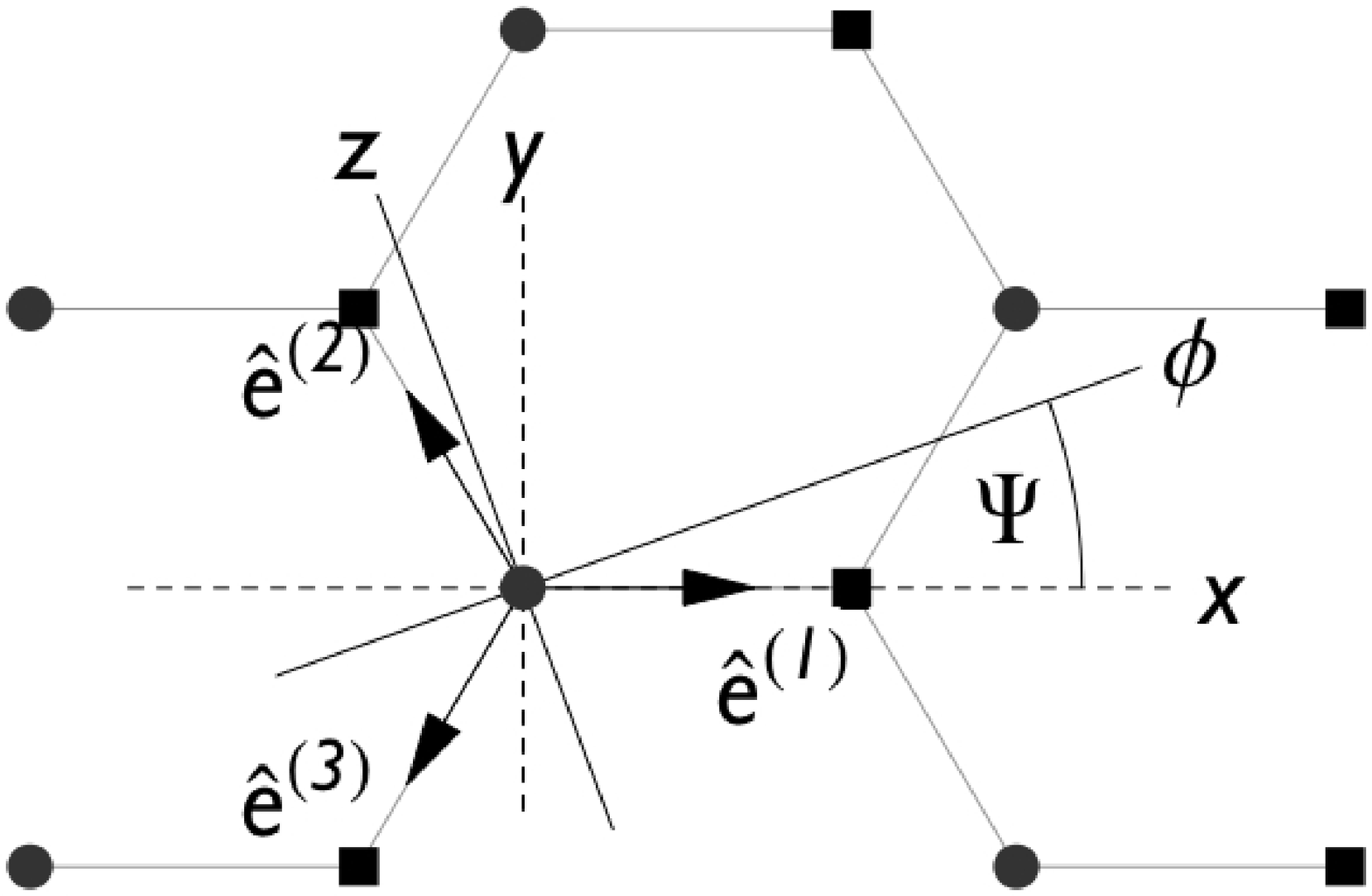}\hspace{3 mm}\epsfxsize 0.8 in 
\epsfbox{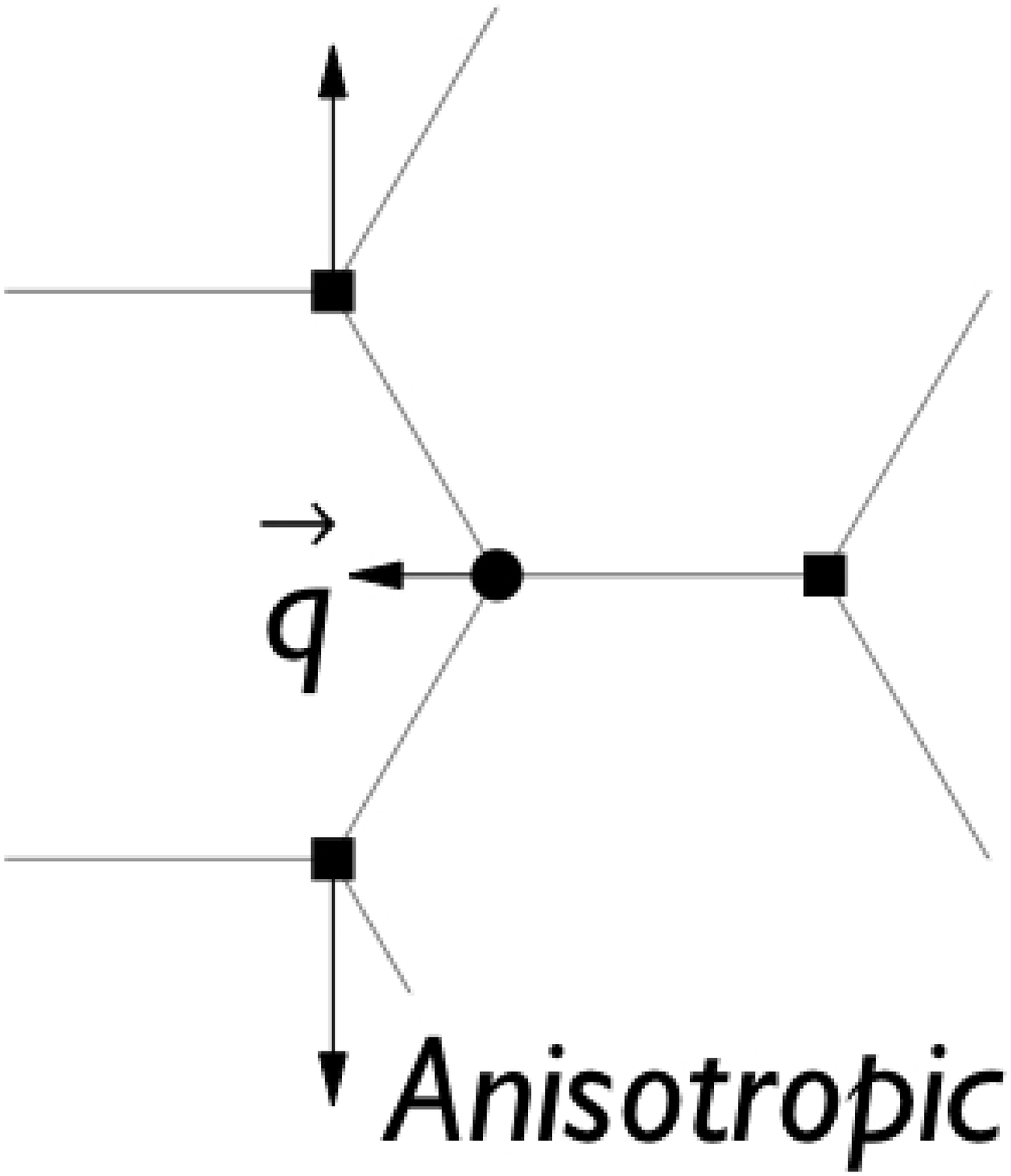}\hspace{3 mm}\epsfxsize 0.8 in
\epsfbox{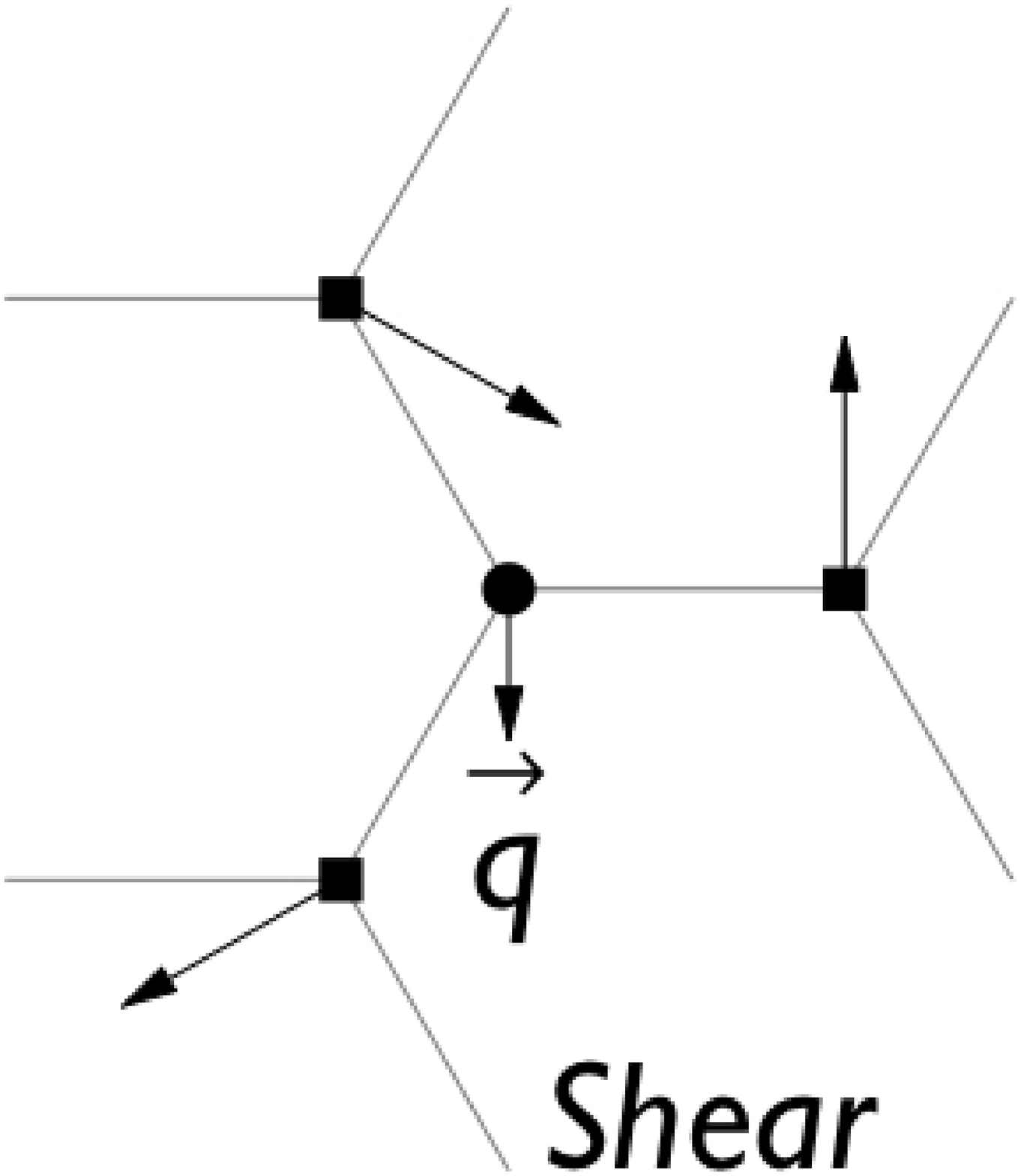}
\vspace{-1 mm}
\caption{The two sublattices (circles and squares) of graphene and the three unit vectors $\hat{e}^{(l)}$ used in the text. ${\phi}$, ${z}$ are cylindrical coordinates of a tube, while $\Psi=\pi/6-\theta_c$ with $\theta_c$ the chiral angle. Also, anisotropic ($u^{xx}=u^{xy}=0$, $u^{yy}=2\!\ \gamma$, $q^x= \ell \!\ \gamma$), shear ($u^{xx}=u^{yy}=0$, $u^{xy}=\eta$,  $q^y=- \ell \!\ \eta$) strains.}
\label{lattice}
\vspace{-2 mm}
\end{figure}

Graphene decomposes into the two triangular sublattices of Fig.~\ref{lattice}. We describe in-plane deformations of the sublattices via two fields, $u^i(x)$, $v^i(x)$, $i=1,2$, and their strain tensors $u^{ij}=\d^{(i}u^{j)} $ and $v^{ij}= \d^{(j}v^{i) }$.  The  density of elastic energy contains direct and cross terms: 
\begin{equation}
V[u,v]= d[u] + d[v] +c[u,v].
\label{V}
\end{equation}
Six-fold symmetry of the sublattices implies isotropy of the direct terms~\cite{Landau}:
\begin{equation} 
d[u]=\mu' \!\ u^{ij}u_{ij}+\frac{\lambda'}{2} \!\ u^i_i u^j_j.
\label{d}
\end{equation} 
Symmetry dictates the  form of the cross term 
\begin{eqnarray}
\label{C}
c[u,v ] &=&2~\mu  u^{ij}v_{ij}+\lambda u^i_i v^j_j    \nonumber \\
           &+&\alpha \(u-v\)^2 \\ \nonumber
           &-&\beta \!\ e_{ijk}\(u^{ij}+v^{ij}\)\(u^k-v^k\)~
           \label{cross}
\end{eqnarray}
The tensor $e_{ijk}$, which is invariant under $C_{3v}$, can be represented by the three unit vectors $\{\hat{e}^{(l)}\}$ of Fig.~\ref{lattice}:
\begin{equation} 
e_{ijk}=\frac{4}{3} \sum_{l=1}^3 \hat{e}^{(l)}_i \hat{e}^{(l)}_j \hat{e}^{(l)}_k. \label{B}
\end{equation} 
Only the last term in Eq.~\ref{cross} is not invariant under general rotation. (In nanotubes, it depends on the helical angle $\theta_c$: $e_{\phi\phi\phi}= - e_{\phi z z}=-\sin (3\theta_c)$, $e_{z z z}= -e_{\phi \phi z}=-\cos (3\theta_c)$, where $\phi$, $z$ are defined in Fig.~\ref{lattice}).  This elastic energy density, the lowest-order  approximation in both derivatives and fields, contains six parameters: $\mu'$ and $\lambda'$, being confined to one sublattice, describe next-neighbor interactions; the cross terms $\mu$ and $\lambda$ describe nearest-neighbor interaction; $\alpha$ describes the stiffness against relative shifts of the sublattices; $\beta$ determines the strength of rotational symmetry breaking and so carries the point group symmetry of graphene. These parameters are normalized to the  sublattice surface density  $\sigma_{s}$, so that the elastic energy is $W=\int \sigma_{s}\!\ V\!\ d^2x$.

Taking $\frac{1}{2}\sigma_{s} \(\dot{u}^2+\dot{v}^2\)$ as the surface density of kinetic energy, the equations of motion read
\begin{equation}
\left\{ \begin{array}{l} 
 \!\ \ddot{u}^i = \d_j \sigma^{ij}_{\(u\)}-2\alpha \(u^i-v^i\)+ \beta\!\ e_{lm}^{~~\!\ i}\!\ \(v^{lm}+ u^{lm}\) \\
 \!\ \ddot{v}^i = \d_j \sigma^{ij}_{\(v\)}+2\alpha \(u^i-v^i\)- \beta \!\ e_{lm}^{~~\!\ i}\!\  \(v^{lm}+ u^{lm}\) 
\end{array} \right.
\label{phonons}
\end{equation}
with  the  sublattice 2-D stress tensors
\begin{equation} 
\left\{\begin{array}{ll}
\sigma^{ij}_{\(u\)} & =  2 \mu' \!\ u^{ij} +\lambda'  \!\ \delta^{ij} u^k_k+ 2 \mu  \!\ v^{ij} +\lambda  \!\ \delta^{ij} v^k_k \\ 
~ &-\beta\!\ e^{ij}_{~~ k}\!\  \(u^k-v^k\) \\
\sigma^{ij}_{\(v\)} & = 2 \mu' \!\ v^{ij} +\lambda' \!\ \delta^{ij} v^k_k+2 \mu \!\ u^{ij} +\lambda \!\ \delta^{ij} u^k_k \\
~& -\beta\!\ e^{ij}_{~~ k}\!\ \(u^k-v^k\) 
\end{array} \right . 
\label{stress}
\end{equation} 
As expected, $\alpha$ determines the frequency of two degenerate $k=0$ optical modes: ${\omega_{\Gamma}}^2 =4\!\ \alpha$.

First, we briefly show that the usual macroscopic elastic energy of graphene and its Lam\'e coefficients can be obtained from $V$ by considering a static, uniform solution of Eqs.~\ref{phonons} with identical deformations on both lattices with an internal displacement $2 q^i\equiv u^i -v^i$ :
\begin{equation} 
2q^i=\ell \!\
e_{lm}^{~~~i}u^{lm}=\ell\!\ e_{lm}^{~~~i}v^{lm} ,
\label{disp}
\end{equation} 
where $\ell =\beta/\alpha$ is a characteristic length. Anisotropic ($2 \gamma= u^{xx}-u^{yy}$) and shear ($\eta= u^{xy}$) strains produce internal displacements $q^x= \ell \!\ \gamma$ and $q^y=- \ell \!\ \eta$ (Fig.~\ref{lattice}). The elastic energy for uniform deformations $W_u=\int V_u \sigma_g \!\ d^2x$  then simplifies  to
\begin{eqnarray} 
V_u[u,q] &=&\(\mu_R+ \frac{\beta ^2}{\alpha}\)\!\
u^{ij}u_{ij} +\frac{1}{2} \(\lambda_R- \frac{\beta ^2}{\alpha}\)\!\ u^i_i u^j_j \nonumber \\
&+& 2 \alpha \!\ q^2 -2 \!\ \beta \!\ e_{ijk}u^{ij}q^k ,
\label{uniform}
\end{eqnarray} 
where  $\sigma_g=2\sigma_s=2.26$  g~cm$^{-2}$ is
the surface density of graphene, $\mu_R \equiv \mu+\mu' -\frac{\!\ \beta ^2}{\alpha}$,
$\lambda_R \equiv \lambda+\lambda'+\frac{\!\ \beta
  ^2}{\alpha}$ the measurable Lam\'e coefficients \cite{Landau}. Macroscopic problems do not distinguish between the two sublatices;  eliminating $q^i$ in Eq.~\ref{uniform} through
Eqs.~\ref{B}~and~\ref{disp} we obtain the familiar, isotropic, macroscopic energy for graphene,
$V_u=\mu_R u^{ij}u_{ij}+ \lambda_R u^i_i u^j_j /2$. In the long wavelength limit
Eqs.~\ref{phonons}  returns the familiar longitudinal and transverse
speeds of sound in terms of the Lam\'e coefficients: $v_{L}^2 =2
\mu_R+\lambda_R$, $v_{T}^2=\mu_R$.

The out-of-plane displacements $u_{\perp}(x)$ and $v_{\perp}(x)$ do not couple with the in-plane $u^i$, $v^i$ in the harmonic limit: invariance under  simultaneous sign change of $u_{\perp}$ and $v_{\perp}$ prevents it, for flat sheets.  Introducing $2 p_{\perp}(x) = u_{\perp}(x)+v_{\perp}(x)$ and $2 q_{\perp}(x) = u_{\perp}(x)-v_{\perp}(x)$,  $V_{\perp}$ must be invariant under $p_{\perp} \rightarrow p_{\perp} +L(x)$, $L(x)$ a linear function in the plane, and  thus, can contain only second (and higher) derivatives in $p_{\perp}$. Symmetry dictates (cf. Appendix)
\begin{eqnarray}
V_{\perp}& =&  4 \alpha_{\perp}q_{\perp}^2 -4 \!\ \alpha'_{\perp} \d_iq_{\perp}\d^i q_{\perp} +4 \beta_{\perp}e_{ijk}\!\ \d^kq_{\perp} \d^{ij}p_{\perp} \nonumber \\
&+& 2 \mu_{\perp}^+ \d_{ij}p_{\perp} \d^{ij}p_{\perp}+\lambda_{\perp}^+\d^i_{i}p_{\perp} \d^i_{i}p_{\perp} \nonumber \\
&-& 2 \mu_{\perp} ^- \d_{ij}q_{\perp} \d^{ij}q_{\perp}-\lambda_{\perp}^-
\d^i_{i}q_{\perp} \d^i_{i}q_{\perp} .
\label{norm}
\end{eqnarray} 
The frequency of the $k=0$ out-of-plane optical mode is $2
\sqrt{\alpha_{\perp}}$, and the out-of-plane acoustic branch is
quadratic at small wave-vector, as expected.

\begin{figure}[t!]
\hspace{-7mm}\epsfxsize 3. in
\epsfbox{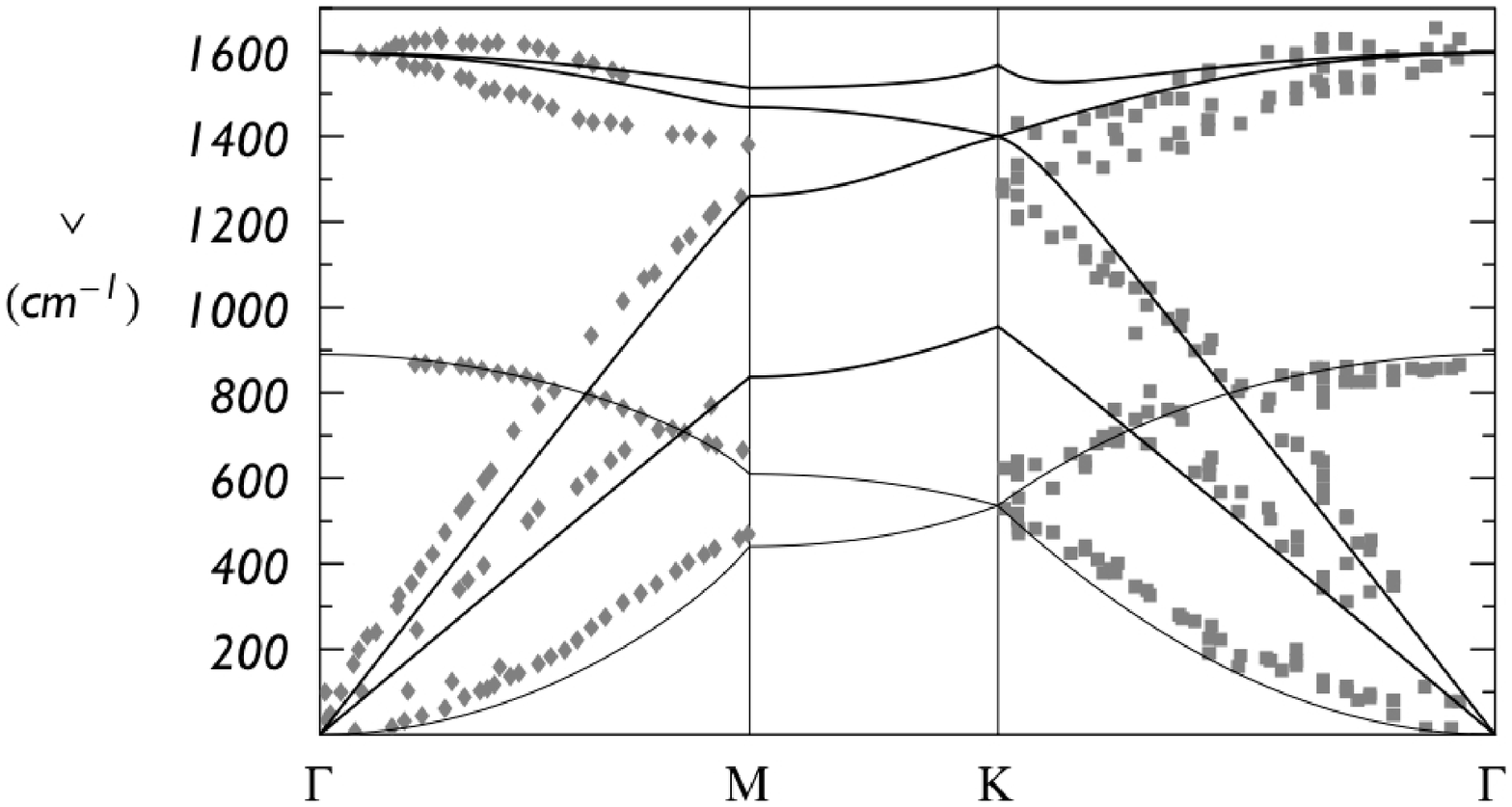}

\hspace{-7mm}\epsfxsize 3. in
\epsfbox{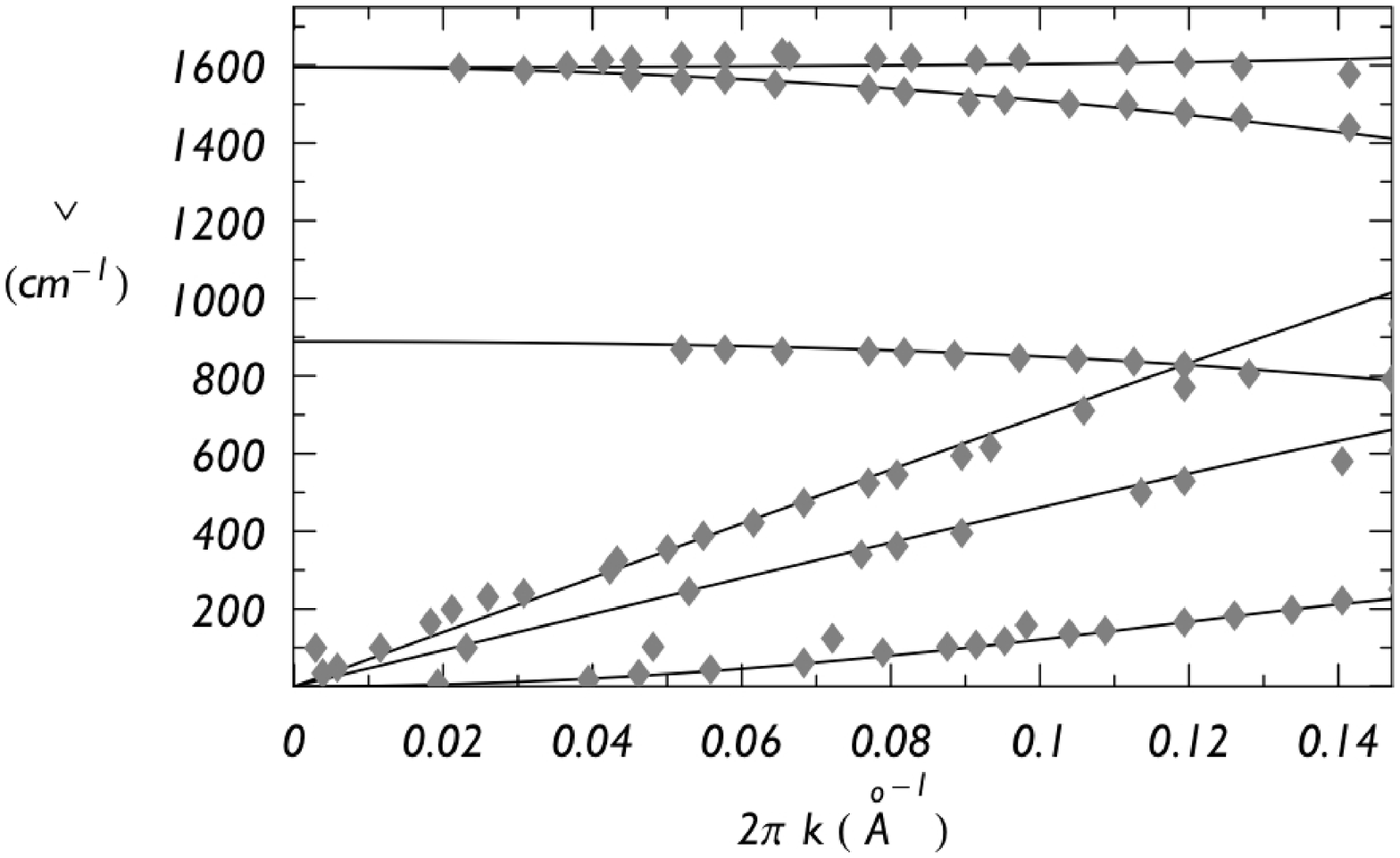}
\caption{Bicontinuum phonons compared to EELS data (diamonds~\cite{Oshima} and squares~\cite{Siebentritt}), fitting either to the entire Brillouin zone (top) or just around $\Gamma$ along $\Gamma \rightarrow M$.}
\vspace{-2 mm}
\label{branches}
\end{figure}

The bicontinuum phonons are much more richly structured than in a traditional continuum model: they include all the optical branches, show nonlinear dispersion at large wavevector, and even display the main features of the Brillouin zone, all without sacrificing the advantages of a continuum framework.  Plane-wave solutions of Eqs.~\ref{phonons} returns an analytically solvable fourth-order secular equation in $\omega(k)$, yielding  two acoustic and two optical branches. The longitudinal branches cross at the vertices of a hexagon. Since the two-field elastic energy density respects the point group symmetry of the graphene lattice, this hexagon is oriented just as the graphene Brillouin zone; 
although the model, unlike in the envelope function approach~\cite{Foreman}, has no built-in length scale, 
the elastic parameters can be constrained so that the crossing point coincides with the $K$ point of graphene. A similar argument holds for the out-of-plane modes: strikingly {\it one can construct the correct Brillouin zone within a continuum model.} Fig.~\ref{branches} shows the bicontinuum phonons fit to electron-energy-loss spectroscopy (EELS) data~\cite{Oshima,Siebentritt} for parameters fitted either to the full Brillouin zone or just around $\Gamma$~\cite{BZ}. 

The bicontinuum provides a unified framework for nanotube mechanics which can describe {\it all} current computational results on the coupling of nanotube phonons to static structural distortions, to each other (e.g. breathing-to-Raman  or longitudinal-to-transverse modes in helical tubes) and to the tube electronic structure. In a cylindrical geometry with coordinates $\{r,\phi, z\}$, a coupling between  the tangential displacements $u^i$, and the radial  $u^r=u_{\perp}$ appears in $V$ of Eq.~\ref{V} via $ u^{\phi\phi}=\(\d_{\phi}u^{\phi}+u^r\)/r$ (and similarly for $v$);  this accounts for the emergence of the Radial Breathing Mode (RBM)~\cite{ccorr}.
We consider uniform solutions: $u=u_o e^{-i \omega t}$, $v=v_o e^{-i \omega t}$.  The tube's helicity can be subsumed into new axes $\{\xi,\zeta\}$ ($\xi={\phi} \cos{3\theta_c}+{z} \sin{3\theta_c}$ $\zeta=-{\phi} \sin{3\theta_c}+{z} \cos{3\theta_c}$) rotated by an angle $3\theta_{c}$ with respect to the base of the tube. In terms of $p$, $q$ we obtain $p^{\xi},p^{\zeta}=0$ and
\begin{equation}
\left\{ \begin{array}{l} 
 q^{\zeta}\(\omega^2 -4\alpha\)+2\frac{\beta}{r}p^r=0\\
 p^r\(\omega^2-\frac{v_L^2+\beta^2/\alpha}{r^2}\) +2\frac{\beta}{r}q^{\zeta}=0  \\
 q^{\xi}\(\omega^2 -4\alpha\)=0\\
 q^r\(\omega^2 -4\alpha_{\perp}+\frac{2\mu-2\mu'+\lambda-\lambda'}{r^2}\)=0\\
\end{array} \right . ~ . 
\label{raman}
\end{equation}
Unlike standard elasticity \cite{Mahan1}, which cannot describe optical modes, or standard atomistic descriptions, which cannot be solved analytically, the two-field continuum model enables an exact analytical solution for the coupling between the RBM and the graphite-like optical mode through the first two of Eqs.~in~(\ref{raman}); the RBM induces a shear in the sublattices, \mbox{$u^{\phi \phi}=v^{\phi \phi}=u^r/r$}, which couples with the internal displacement through $\beta$, and vice versa. Thus, the RBM is not purely radial, but has a longitudinal component  $q_B^z\sim
\frac{\ell}{2 r}\cos{3\theta_c}$, as previously seen in a numerical calculation\cite{serbians}. Expansion of the RBM frequency in powers of $l/r$ reveals a correction to the the standard continuum result $v_L/r$~\cite{Mahan1}:
$\omega_B=\frac{v_L}{r}\left[1-\frac{1}{8} \(\frac{\ell }{r}\)^2+O\(\frac{\ell }{r}\)^4\right]$. The graphite-like optical modes of chiral tubes are $\omega_{\xi}=\sqrt{4\,\alpha}$, $\omega_{\zeta}/\omega_{\xi}=1+\frac{1}{8} \(\frac{l}{r}\)^2+O\(\frac{l}{r}\)^4$, also of mixed longitudinal/transverse character except for armchair and zig-zag nanotubes, while the out-of-plane optical mode $\omega_{\perp}= \left(4\alpha_{\perp}-\frac{2\mu-2\mu'+\lambda-\lambda'}{r^2}\right)^{1/2}$ is purely radial. 
A density functional theory calculation of the breathing mode~\cite{breathing} reports different frequencies with ($\omega_B$) and without ($\tilde \omega_{B}$) coupling to optical modes. We predict  $r^2 \(\tilde\omega_{B}^2-\omega_{B}^2\)\rightarrow \beta^2/\alpha$ as $r\rightarrow \infty$: using ref~\cite{breathing} data for $\tilde\omega_{B}$, $\omega_{B}$ we obtain $\ell \equiv \beta/\alpha=0.25$ \AA~ ($0.27$ \AA) for non metallic zig-zag (armchair) tubes, in good agreement with the parameters from our fit to the graphene phonons~\cite{BZ}. 

The bicontinuum can also describe electron-lattice coupling to both acoustic and optical modes, by incorporating a tight-binding model whose  nearest neighbor hopping integrals  $t^{(1)},t^{(2)},t^{(3)}$ are modulated by the in-plane elastic deformations:
\begin{equation} 
\mathrm{d}t^{(l)}= - \tau \!\ \hat{e}^{(l)}_i \hat{e}^{(l)}_j u^{ij} +\tau \!\ \hat{e}^{(l)}_i q^i/e
\label{hopping}
\end{equation}  
where $e$ is the inter-atomic distance and $\tau$ a parameter to be determined~\cite{hopping}. For example, lattice deformations open gaps in metallic tubes, and these gaps in turn affect vibrational frequencies. If $\epsilon_{c}$, $\epsilon_{v}$ are the conduction and valence bands, we have to nearest neighbors 
\begin{equation} 
\epsilon_{c}(k)^2-\epsilon_{v}(k)^2 =\sum_{l}   t^{(l)}+2\sum_{m>l} t^{(l)} t^{(m)}\cos(k\cdot a^{(n)}),
\label{band}
\end{equation} 
where $a^{(n)} \equiv e^{(l)}-e^{(m)}$, $n(l, m)$ is  cyclic  in $\{1,2,3\}$ (e.g.  $a^{(3)}
\equiv e^{(1)}-e^{(2)} $) and $\{e^{(i)}\}$ connects nearest neighbors. 
From Eqs. \ref{hopping},\ref{band} we find the band gap opened by strain in a  metallic nanotube to be
\begin{eqnarray}
\frac{\Delta^2}{\(3  \tau\)^2}&=&\frac{1}{2} u^{ij} u_{ij}- \frac{1}{4}u_i^i u_i^i - \frac{1}{e}\!\  e_{ijk}u^{ij} q^k\nonumber \\
&+& \frac{1}{e^2} \!\ \(\hat{z}_i q^i\)^2 +\frac{1}{e}\!\ e_{ijk} \hat{\phi}^k u^{ij} \hat{\phi}_h q^h-\frac{1}{4} \(e_{ijk}u^{ij} \hat{\phi}^k\)^2. \nonumber \\
\label{gap}
\end{eqnarray} 
In the second line of equation (\ref{gap}) the symmetry of the honeycomb lattice is broken by the unit vectors  $\hat{\phi}^i, \hat{z}^i$ of the cylindrical coordinates. In terms of $2 \gamma' \equiv u^{\phi
  \phi}-u^{z z}$, $\eta' \equiv u^{\phi z}$, $q^z$, equation (\ref{gap}) reads
\begin{equation} \Delta=3 \tau\left |q^z/e+\gamma'
    \cos(3\theta_c) +\eta' \sin(3\theta_c) \right |,
    \label{Yang-nisoli}
\end{equation} 
which corrects and extends a well known previous result within a one-field continuum model~\cite{Yang} that neglected the inner displacement (i.e. $q^i=0$). 

Opening bandgaps in metallic nanotubes causes several shifts in observed quantities.  The term proportional to $q_z^2$ in Eq. (\ref{gap}) show that longitudinal optical modes open a bandgap in metallic tubes of any helicity;  the elastic energy lowers by a term proportional to the square of the bandgap, leading to a the softening of  longitudinal optical frequency in metallic nanotubes, as revealed by a recent DFT study~\cite{Dubay}.  Eq. (\ref{gap}) predicts also a softening of the RBM in metallic nanotubes $\frac{\delta \omega_B}{\omega_B}=-A \cos^2\(3\theta_c\)$, highest for zig-zag tubes
as seen in DFT~\cite{breathing}, and relates it to the optical softening,  with $A=(1-\ell/e) \omega_{opt}\delta \omega_{opt} e^2/4 v_L^2$, $\omega_{opt}$ the graphite-like optical mode, and $\delta \omega_{opt}$ its softening in metallic tubes ($A\simeq2\%$). Other shifts can be predicted: 
 the speed of  sound for the twist mode softens by $\frac{\Delta c_t}{ c_t}=- \frac{v_L^2}{2 v_T^2}A \sin^2 \(3\theta_c\)$, or $\simeq 2.2\%$ in armchair tubes.
 
Doping-induced structural deformations can also be studied by minimizing the total energy (elastic plus doped electrons). Subtle phenomena absent in other models~\cite{Verissimo} can be accessed within the bicontinuum framework. Going to next-nearest-neighbor in the hopping integrals ($\mathrm{d}t_1^{(l)}= - \tau_1 \!\ \hat{a}^{(l)}_i \hat{a}^{(l)}_j v^{ij}$~\cite{hopping}), we find that  at first order in both $a/r$ and the number of dopant electrons per atom $\rho_e$, semiconducting $(n,0)$ nanotubes show doping-induced changes in tube length ($\mathrm{d} L/L=u^{z z}$) and axial bond-length ($\mathrm{d}b_{ax}=e u^{zz}-q^z$):
\begin{equation}
\left\{ \begin{array}{l} 
\mathrm{d} L/L=\frac{\rho_e\tau}{8 m_C v_T^2 }\left[\pm\(1-\frac{\ell}{e}\)+ \frac{3 \tau_1}{2 \tau}\frac{2\mu_R+\lambda_R}{\mu_R+\lambda_R}\right] \\
\mathrm{d}b_{ax}=\pm\frac{\rho_e \tau}{2 m_C \omega_{opt}^2 e }\\
\end{array} \right . ~ . 
\label{roxana}
\end{equation}
where $m_C$ is the mass of the carbon atom.  The sign is positive (negative) for $r=n\mod3 = 2$ ($n\mod3 = 1$). Recent DFT results~\cite{Margine} indeed  show shrinking or stretching of $b_{ax}$ for $n=16,13$ or $n=14,11$ tubes respectively, as predicted by Eq.~\ref{roxana}. In DFT, the overall tube lengthens in the second case ($n=14,11$), again in accord with the bicontinuum; the lengthening found for $r=2$, is less than for $r=1$, perhaps a consequence of the change in sign in Eqs.~\ref{roxana}. Finally the shrinking of the axial bond determines an up-shift in the longitudinal graphite-like optical mode and might explain recent Raman results that point toward anomalous bond contraction under doping in semiconducting nanotubes~\cite{Chen1,Chen2}.

In summary, a symmetrized two-field continuum model of graphene and carbon nanotubes provides the first unified analytical treatment for a wide range of vibrational and electromechanical phenomena including nonlinear dispersion of in-plane phonons, zone-edge degeneracies and optical modes. A full range of vibrational-electronic-mechanical couplings, which were absent from previous continuum models or happened upon in an ad hoc fashion in computational work, can now be understood within a single unified analytical framework. Extending the formalism to include higher-order effects arising from curvature or metallic character (i.e. symmetry breaking terms containing $\hat{\phi}^i$, $\hat{z}^i$, as in Eq.~\ref{gap}), anharmonicity (terms higher order in $u^{ij}$, $v^{ij}$), or long-distance interactions (higher partial derivatives) is straightforward.  An extension to boron nitride nanotubes, with different coefficients for each sublattice in the direct terms of Eq.~\ref{d}, might prove useful to study their piezoelectricity.

\vspace{-5mm}
\subsection{Appendix: Derivation of Eq.~\ref{C}}
\vspace{-4mm}

The term $c[u,v]$ must be invariant under the combination of $2\pi/6$ rotations and the exchange of fields $u\leftrightarrow v$. Adding reflection through the $x$ axis (Fig.~\ref{lattice}) then implies $C_{3v} $ invariance. There is also a field translation invariance: $u(x)\rightarrow u(x)+p$, $v(x)\rightarrow v(x)+p$.  The objects $u^i$, $v^j$, $u^{ij}$, and $v^{ij}$ can be combined pairwise only into tensors of rank two, three and four; thus $c[u,v]$ decomposes into three parts. The first part has terms like $u^i v^j$; symmetry then implies the form $\alpha (u-v)^2$ with $\alpha > 0$ to ensure an energy minimum. The second part has terms like $u^{ij} v^{kl}$; the only admissible form is $2 \mu u^{ij}v_{ij}+\lambda u^i_i v^j_j$. The third part contains only rank three terms such as $u^{ij}v^k$ contracted with a $C_{3v}$ invariant tensor $e_{ijk}$, giving $e_{ijk} u^{ij}v^k$. By requiring invariance under $2\pi/6$ rotations conjugated with sublattice switching, and also the field translation invariance, we obtain the form $e_{ijk}
u^{ij}(u^k-v^k)+e^*_{ijk} v^{ij}(v^k-u^k)$, where the star means a $2\pi/6$ rotation. Since $C_{3v}$ invariance implies $e^*_{ijk}=-e_{ijk}$ we finally obtain the third row of Eq.~\ref{C}.


\begin{thebibliography}{}
\bibitem{BZ}  The fit around $\Gamma$ returns (in Km~s$^{-1}$) $v_{L}=21$, $v_{T}=14$, $\(-2\mu+2\mu'-\lambda+\lambda'+\beta^2/\alpha\)^{1/2}=4.4$,  $\(\mu-\mu'\)^{1/2}=15,6$ and $\ell \equiv \beta/\alpha=0.3$ \AA~\cite{BZ}. The fit to the full zone uses $v_{L}=16.5$ Km s$^{-1}$, $v_{T}=10.8$ Km s$^{-1}$, $\(2\mu-2\mu'+\lambda-\lambda'-\beta^2/\alpha\)^{1/2}=8.7$ Km s$^{-1}$,  $\(\mu-\mu'\)^{1/2}=6,6$ Km s$^{-1}$, $\ell \equiv \beta/\alpha=0.24$ \AA; an extension to higher derivatives would improve the agreement.
\bibitem{ccorr} The chiral vector of the tube breaks the hexagonal symmetry and allows for new terms to be introduced in $V$ as curvature corrections, which for simplicity we won't consider here. Different problems will suggest different leading corrections.
\bibitem{hopping} $t^{(l)}$, $t_1^{(l)}$ are the absolute values of the hopping integrals for nearest and next nearest neighbors. We assume they depend only on distance, and thus $\mathrm{d}t=-\tau \mathrm{d}e/e$,  $\mathrm{d}t_1=-\tau_1 \mathrm{d}a/a$.
\end{thebibliography}
\end{document}